\documentclass{article}
\usepackage[all]{xy}
\usepackage{authblk}
\usepackage{amssymb}
\usepackage{amsmath}

\def\beq{\begin{equation}}
\def\eeq{\end{equation}}
\def\({\left(} \def\){\right)}
\def\na{\nabla}
\def\nab#1{{\buildrel #1\over \na}}
\def\sFrac[#1/#2]{\hbox{$\frac{#1}{#2}$}}
\def\Frac[#1/#2]{\frac{#1}{#2}}

\def\calR{{\cal R}}

\def\Lor{\hbox{Lor}}
\def\Con{\hbox{Con}}
\def\Mat{\hbox{Mat}}
\def\Jor{\hbox{Jor}}
\def\Hel{\hbox{Hel}}
\def\Ein{\hbox{Ein}}
\def\BD{\hbox{BD}}
\def\R{{\mathbb R}}
\def\arr{\rightarrow}

\def\Ga{\Gamma}
\def\al{\alpha}
\def\be{\beta}
\def\de{\delta}

\def\ka{\kappa}
\def\ep{\epsilon}
\def\vp{\varphi}

\begin{document}

\title{Extended Gravity in EPS Formalism}
\author[a,b,c]{S.Capozziello}
\author[d,e]{L.Fatibene}
\author[d,e]{S.Garruto}

\affil[a]{Dipartimento di Fisica, University of Napoli}
\affil[b]{INFN Sezione  Napoli -- Iniziativa Specifica QGSKY}
\affil[c]{ Gran Sasso Science Institute (INFN), L'Aquila (Italy) }
\affil[d]{Dipartimento di Matematica, University of Torino (Italy)}
\affil[e]{INFN Sezione  Torino -- Iniziativa Specifica QGSKY}

\date{}

\maketitle

\begin{abstract}
We shall discuss equivalence of frames in Palatini $f(\calR)$-theories at action level.
A Palatini formulation of Brans-Dicke theories (equivalent to the purely metric ones) will also be discussed.
\end{abstract}



\section{Introduction}

Integrable Extended Theories of Gravitation (iETG) are Palatini field theories for a metric $g$ and a (torsionless) connection $\tilde \Ga$
(as well as, possibly, matter fields) in which the field equations enforce the compatibility condition 
\beq
\tilde \Ga^\al_{\be\mu} = \{g\}^\al_{\be\mu} - \sFrac[1/2]\(g^{\al\ep} g_{\be\mu} -2 \de^\al_{(\be}\de^\ep_{\mu)}\)\nab{\ast}_\ep \ln \vp 
=\{\vp g\}^\al_{\be\mu} 
\label{CC}
\eeq
which implements the Ehlers-Pirani-Schild (EPS) framework for gravitation and spacetime geometry; see \cite{EPS,DiMauro,Extended2,EG,EC, Cap1, Cap2,Faraoni,DeFelice}.
Here and hereafter the symbol $\nab{\ast}_\ep$ denotes the covariant derivative when it does not depend on any connection (e.g.~when applied to scalar fields), $\na_\ep$ will denote covariant derivative with respect to the original metric $g$,
$\tilde\na_\ep$ will denote covariant derivative with respect to the conformal metric $\tilde g=\vp g$, or more generally with respect to the connection $\tilde \Ga$.

The Palatini $f(\calR)$--theories are particular iETG.
In this paper we shall propose a definition of {\it frame}, equivalence of frames, and provide a number of examples of different frames in the context of  Palatini $f(\calR)$--theories.

Let us define a {\it variational frame} to be a Lagrangian with a specific choice of fundamental fields which one can vary independently.
Two variational frames are said to be {\it dynamically equivalent} if there is a 1-to-1 correspondence between solutions of their field equations.
A {\it variational principle} is an equivalence class of dynamically equivalent variational frames.

The {\it Jordan frame} for a  Palatini $f(\calR)$--theory is defined to be the choice of the fundamental fields $(g_{\mu\nu}, \tilde \Ga^\al_{\be\mu}, \psi)$ for the Lagrangian
\beq
L_J= \sFrac[1/2\ka] \sqrt{g} f(\calR)  + L_m(g, \psi)
\eeq
where $\sqrt{g}$ denotes the square root of the absolute value of the determinant of the metric $g$, $\calR$ is defined to be $\calR= g^{\mu\nu} \tilde R_{\mu\nu}(\tilde \Ga)$, and $\psi$ is a collection of matter fields. The function $f(\calR)$ is a sufficiently regular function (fixed once and for all so that the master equation defined below can be solved for the curvature).

Field equations in the Jordan frames read as 
\beq
\begin{cases}
f'(\calR) \tilde R_{(\mu\nu)}-\sFrac[1/2] f(\calR) g_{\mu\nu}= k T_{\mu\nu}\\
\tilde \na_\mu \( \sqrt{g} f'(\calR) g^{\al\be}\)=0\\
E=0
\end{cases}
\eeq
where $E=0$ denote matter field equations and $T_{\mu\nu}$ the energy--momentum stress tensor obtained as a variational derivative of the matter Lagrangian with respect to the metric $g$. We assumed that  matter does not couple to the connection but to the metric only.
The second field equation implies the compatibility condition (\ref{CC}) so that the connection $\tilde \Ga$ turns out to be the Levi Civita connection of a conformal metric $\tilde g= \vp \cdot g$, where the conformal factor has been set to $\vp=f'(\calR)$ (in 4d).

Another  important consequence of the field equations is the trace (by $g$) of the first field equation, namely (in 4d)
\beq
f'(\calR) \calR -2 f(\calR) = k T
\eeq
where we set $T= g^{\mu\nu} T_{\mu\nu}$.
This is called the {\it master equation}. It is an algebraic (not differential) equation and $f(\calR)$ is chosen so that one can solve (at least locally) for the curvature $\calR=\calR(T)$, and it encodes a lot of properties of the theory; see \cite{Univ, EC}. 

For a given function $f(\calR)$ one can write the equation of the conformal factor $\vp= f'(\calR)$ and solve it for the curvature
$\calR= r(\vp)$. If the function $f'(\calR)$ is 1-to-1 then the function $r$ is global (otherwise one can invert single branches to obtain a number of local inverses). 
For each of such a pair of functions $(f(\calR), r(\vp))$ one can define the following Lagrangian
\beq
L_H= \sqrt{g}\( \vp \calR +U(\vp)\) + L_m(g, \psi)
\eeq
where we set $U(\vp)=\vp\( f(r(\vp))-r(\vp)\)$ for the potential.
The choice of independent fields $(g, \tilde \Ga, \vp, \psi)$ is called the {\it Helmholtz frame}.

Field equations in the Helmoltz frame can be recast as 
\beq
\begin{cases}
f'(\calR) \tilde R_{(\mu\nu)}-\sFrac[1/2] f(\calR) g_{\mu\nu}= k T_{\mu\nu}\\
\tilde \na_\mu \( \sqrt{g} f'(\calR) g^{\al\be}\)=0\\
\end{cases}
\qquad
\begin{cases}
\calR- r(\vp)=0\\
E=0\\
\end{cases}
\eeq
Accordingly, if the function $f'(\calR)$  is invertible, then Helmoltz frame and Jordan frames are dynamically equivalent by the map
$\vp=f'(\calR)$ and they define the same variational principle.
If the function $f'(\calR)$  is not invertible, then one has a number of branches $r_i(\vp)$ and can define a  Helmoltz frame on each branch and Jordan frame is dynamically equivalent to the union of all Helmoltz frame defined in this way.

\section{Bundle setting for frames}

Let $\Lor(M)$ be the bundle of Lorentzian metrics, $\Con(M)$ the bundle for (torsionless) connection on spacetime $M$,  
$S= M\times \R$ the bundle for a scalar field on $M$, and $\Mat$ the bundle for matter fields.
The function $\hat h: \calR \mapsto \vp= f'(\calR)$ in fact defines a bundle map 
\beq
\begin{aligned}
h:& \Lor(M)\times j^1 \Con(M) \times j^1\Mat \arr \Lor(M)\times j^1\Con(M) \times S\times j^1\Mat\\
  :& (g_{\mu\nu}, j^1\tilde \Ga^\al_{\be\mu}, j^1\psi) \mapsto  (g_{\mu\nu}, j^1\tilde \Ga^\al_{\be\mu}, \vp= f'(\calR), j^1\psi)
\end{aligned}
\eeq
where $j^1$ denotes the first jet prolongation (see \cite{Saunders, Book}) and the Lagrangian in the Jordan frame $L_J= h^{\ast}(L_H)$ is obtained by pull-back from the Helmoltz Lagrangian.

This bundle map also induces a 1-to-1 correspondence between sections of $\Jor=Lor(M)\times \Con(M) \times \Mat$
and sections of $\Hel=\Lor(M)\times \Con(M) \times S\times \Mat$ which restricts to the map for dynamical equivalence. 
Notice that we are not defining any bundle map $h: \Jor\arr \Hel$, only a map between sections which comes from a map between some jet prolongation.

Another, somehow simpler, example is the Levi Civita bundle map
\beq
LC: j^1\Lor(M) \arr \Lor(M) \times \Con(M): j^1 g_{\mu\nu}\mapsto (g_{\mu\nu}, \{g\}^\al_{\be\mu})
\eeq
which is prolonged to a map $jLC: j^2\Lor(M) \arr \Lor(M) \times j^1\Con(M)$.
Any Palatini Lagrangian on $\Lor(M) \times j^1\Con(M)$ induces a purely metric Lagrangian on $j^2\Lor(M)$; 
however, the map induced on sections is not invertible and only some dynamics (e.g.~standard GR) are seen to be dynamically equivalent; see \cite{B}.

\section{Einstein frame}

One can consider the map $\hat E: \Lor(M)\times S \arr \Lor(M) \times S: (g, \vp)\mapsto (\tilde g= \vp \cdot g, \vp)$ and the bundle
$\Ein= \Lor(M)\times \Con(M)\times S\times \Mat$.

Let us first define a frame $(\tilde g, \tilde \Ga, \vp, \psi)$.
The map $E$ defined on sections is 1-to-1 and the Helmoltz Lagrangian induces a dynamically equivalent Lagrangian
\beq
L_E= \sqrt{\tilde g} \( \tilde R + U(\vp)\)+ L^\ast_m(\vp,  \tilde g, \psi)
\eeq
where we set $\tilde R= \tilde g^{\mu\nu} \tilde R_{\mu\nu}(\tilde \Ga)$ and $L^\ast_m(\vp,\tilde g,  \psi)=L_m(\vp^{-1} \tilde g, \psi)$.
This is a standard GR (for the conformal pair $(\tilde g, \tilde \Ga)$) with the conformal factor $\vp$ which acts as an extra matter field and a modified matter Lagrangian to account for the extra coupling with the conformal factor.

If the matter Lagrangian $L_m$ is conformally invariant (as, e.g., Maxwell Lagrangian in 4d) then 
$L^\ast(\vp,\tilde g,  \psi)= L_m(\tilde g, \psi)$ and one has Maxwell Lagrangian in the new frame as well.
If not, as it happens for example with a Klein-Gordon field, one can extend the conformal transformation to act on $\psi$ as
\beq
\hat E: \Lor(M)\times S\times S \arr \Lor(M)\times S\times S:(g, \vp, \psi)\mapsto (\tilde g, \vp, \tilde \psi=\vp^{-1/2} \psi)
\eeq
which extends to a 1-to-1 bundle map $E: \Hel \arr \Ein$.
The frame $(\tilde g, \tilde \Ga, \vp,\tilde\psi)$ is called the {\it Einstein frame} and the Lagrangian reads as
\beq
L_E= \sqrt{\tilde g}  \tilde R + \tilde L_m(\vp,  \tilde g, \tilde \psi)
\eeq
where we set $\tilde L_m(\vp,  \tilde g, \tilde \psi)= L_m(\vp^{-1}\tilde g,  \vp^{1/2} \tilde\psi) + \sqrt{\tilde g}  U(\vp)\vp^{-2}$.

If one sets the matter Lagrangian to represent a Klein-Gordon field in the Jordan frame, in the Einstein frame one has
\beq
 \tilde L_m(\vp,  \tilde g, \tilde \psi)= -\sFrac[\sqrt{\tilde g}/2] \( \tilde g^{\mu\nu} \nab{\ast}_\mu \tilde \psi \tilde \na_{\nu} \tilde\psi +  \tilde m^2 \tilde\psi^2\)
\eeq
where one has
\beq
\tilde m^2= \sFrac[ m^2/\vp] +\sFrac[5/4\vp^2] \na_\mu\vp \tilde \na^\mu \vp - \sFrac[1/\vp] \Box\vp 
\eeq

This is not a free Klein-Gordon field (since the mass is not constant it has a coupling with the conformal factor).
However, for a conformal factor which is almost constant, namely $\vp= \vp_0 + \ep \de \vp$, the new field can be  considered as a
(minimally coupled) Klein-Gordon field (with a weak interaction with the conformal factor variation) and a mass of $\tilde m^2= \vp_0^{-1} m^2$.
Accordingly, the conformal transformation maps Klein-Gordon fields into (different) Klein-Gordon fields.

\section{Brans-Dicke frame}

Another field transformation one can do is defining a new connection 
\beq
\Ga^\al_{\be\mu}= \tilde \Ga^\al_{\be\mu} + \sFrac[1/2]\(g^{\al\ep}g_{\be\mu} -2 \de^\al_{(\be}\de^\ep_{\mu)}\)\nab{\ast}_\ep\ln\vp
\label{HBD}
\eeq
This is a map from $\Lor(M)\times \Con(M)\times j^1S\times \Mat$ to the bundle $\BD= \Lor(M)\times \Con(M)\times S\times \Mat$.
Then the variatioanl frame $(g_{\mu\nu}, \Ga^\al_{\be\mu}, \vp, \psi)$ is called the {\it (metric-affine) Brans-Dicke frame}.

Dynamics in the BD frame is described by a Palatini Lagrangian
\beq
L_{BD}(g, \Ga, \vp, \psi) = L_H(g, \tilde\Ga( \Ga, g, j^1\vp), \vp, \psi)
\eeq
Since the transformation (\ref{HBD}) is affine in the connection, then the Helmoltz frame and the BD frame are dynamically equivalent.

By setting the connection $\Ga=\{g\}$ one defines the {\it purely metric Brans-Dicke frame} $(g, \vp, \psi)$ in which one recovers the stadard Brans-Dicke theories.

\section{Conclusions and perspectives}

We defined frames and discussed dynamical equivalence at the level of action.
Only at action level one can motivate the name for the Einstein frame since also in the Jordan frame (as whenever field equations depend linearly on the Ricci tensor of some conformal metric) one can recast equations in Einstein form; see \cite{Bb, Bc}.

The (metric-affine) BD frame is, to the best of our knowledge, new. In the Lagrangian $L_{BD}$ there are a number of contributions 
in $\Ga-\{g\}$ which disappear when one goes to the metric BD frame.
While (metric-affine) BD frame and purely metric BD frames are dynamically equivalent, the dynamical equivalence is lost if one just promotes the connection in the purely metric BD frame to be an independent field. This is just because doing so, the terms in  $\Ga-\{g\}$ are not restored.

The frame equivalences we discussed can be summarised by the following diagram:
\medskip 
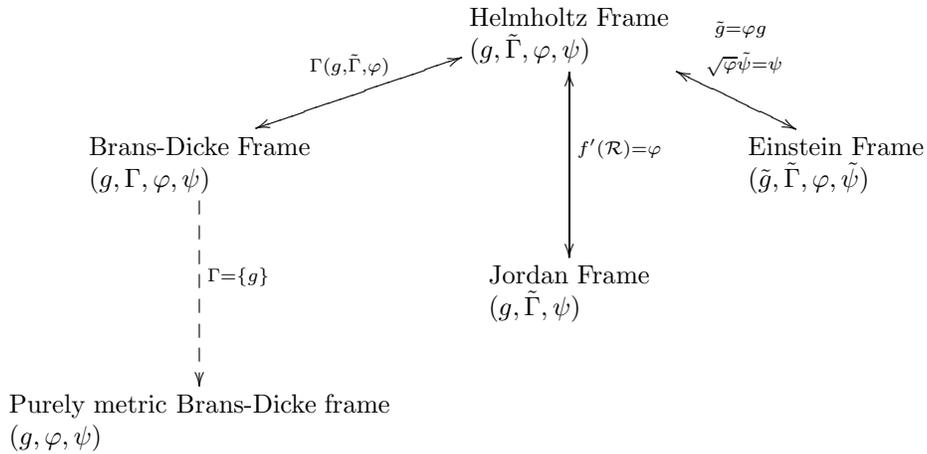
\def\HelBox{\vtop{\hbox{Helmholtz Frame}\hbox{$(g, \tilde \Ga, \vp, \psi)$}}}
\def\JorBox{\vtop{\hbox{Jordan Frame}\hbox{$(g, \tilde \Ga,  \psi)$}}}
\def\BDBox{\vtop{\hbox{Brans-Dicke Frame}\hbox{$(g, \Ga,  \vp, \psi)$}}}
\def\EinBox{\vtop{\hbox{Einstein Frame}\hbox{$(\tilde g, \tilde \Ga, \vp,\tilde\psi)$}}}
\def\BDmBox{\vtop{\hbox{Purely metric Brans-Dicke frame}\hbox{$( g,\vp, \psi)$}}}
\def\LabBox{\vtop{\hbox{ $_{{\tilde g=\vp g}}$}\hbox{$_{{\sqrt{\vp}\tilde \psi= \psi}}$}}}
\begin{equation*}
\xymatrix{ 
		&	\HelBox\ar@{<->}[dd]^{{f'(\calR)=\vp}}  \ar@{<->}[dl]_{{\Ga(g, \tilde \Ga, \vp)\kern10pt}}    \ar@{<->}[dr]^{\kern10pt\LabBox}&     \\
\BDBox\ar@{-->}[dd]^{{\Ga=\{g\}}}		&	 &    \EinBox \\
		&	\JorBox &     \\
\BDmBox		&	 &     \\
 }
\end{equation*}
\medskip

Future investigations will be devoted to consider conformal transformations on Dirac fields and to discuss how  matter in the Einstein frame behaves. In fact, in applications to cosmology one can expect the conformal factor to be slowly dependent on time so that the mass of the Klein-Gorgon field $\tilde \psi$ is expected to vary during the evolution of the universe.
This variation is quite constrained by spectral lines which are observed to be deformed only by cosmological redshift.

Preliminary results show that when the field $\tilde\psi$ is considered to couple with $g$ all the affects of the conformal factor (except the ones associated to its variation which is assumed small) can be hidden in a redefinition of an effective Newton constant.
Accordingly, it does not affect quantum physics at the standard scale.

\section*{Acknowledgments}
We are grateful to A.Borowiec for comments and discussion.
We acknowledge the contribution of INFN (Iniziativa Specifica QGSKY), 
the local research project {\it Metodi Geometrici in Fisica Matematica e Applicazioni} (2015) of Dipartimento di Matematica of University of Torino (Italy). 
This paper is also supported by INdAM-GNFM.

\end{document}